\documentstyle[psfig]{l-aa}     

\def\kms{km~s$^{-1}$ }
\def\a85{ABCG~85{}}
\def\bj{B$_{\rm J}$ }
\thesaurus{11.03.4 ABCG~85; 11.03.1 }

\title{A photometric catalogue of galaxies in the cluster Abell 85.
\thanks{Based on plates scanned with the MAMA microdensitometer at CAI, Paris
and on observations collected at the European Southern Observatory, La Silla,
Chile}
\thanks{Tables 1 and 2 are only available in electronic form at the CDS via
anonymopus ftp to cdsarc.u-strasbg.fr (130.79.128.5) or via
http://cdsweb.u-strasbg.fr/Abstract.html}
}

 \author {
  E.~Slezak \inst{1}
\and
  F.~Durret \inst{2,3}
\and
  J.~Guibert \inst{4}
\and
  C.~Lobo \inst{2,5}
}
\offprints{E.~Slezak, slezak@obs-nice.fr}
\institute{
    Observatoire de la C\^ote d'Azur, B.P. 229, F-06304 Nice Cedex 4, France 
\and
  Institut d'Astrophysique de Paris, CNRS, Universit\'e Pierre et Marie Curie, 
  98bis Bd Arago, F-75014 Paris, France 
\and 
    DAEC, Observatoire de Paris, Universit\'e Paris VII, CNRS (UA 173),
    F-92195 Meudon Cedex, France 
\and
    CAI et Observatoire de Paris, 61 Avenue de l'Observatoire, F-75014 Paris,
    France
\and
    Centro de Astrof\'\i sica da Universidade do Porto, Rua do Campo Alegre 
    823, P-4150 Porto, Portugal 
}
\date{Received, 1997; accepted,}
\begin{document}

\maketitle

\begin{abstract}
We present two catalogues of galaxies in the direction of the rich cluster \a85.
The first one includes 4,232 galaxies located in a region $\pm 1^\circ$
from the cluster centre. It has been obtained from a list of more than
25,000 galaxy candidates detected by scanning a Schmidt photographic plate
taken in the \bj band.
Positions are very accurate in this catalogue but magnitudes are not.
This led us to perform CCD imaging observations in the V and R bands to
calibrate these photographic magnitudes. A second catalogue (805
galaxies) gives a list of galaxies with CCD magnitudes in the V and R bands
for a much smaller region in the centre of the cluster. \\
These two catalogues will be combined with 
a redshift catalogue of 509 galaxies (Durret
et al. 1997) to investigate the cluster properties at optical wavelengths 
(Durret et al. in preparation), as a complement to our previous X-ray studies
(Pislar et al. 1997, Lima--Neto et al. 1997).\\
\end{abstract}

\section{Introduction}

\a85 is a very rich cluster located at a redshift z=0.0555. We
performed a detailed analysis of this cluster from the X-ray point of
view, based on Einstein IPC data (Gerbal et al. 1992 and references
therein).  In the optical, no photometric data were available at that
time, except for an incomplete photometric catalogue by Murphy (1984),
and about 150 redshifts were published in the literature only after we
completed our first X-ray analysis (Beers et al. 1991, Malumuth et
al. 1992). We therefore undertook a more complete analysis of this cluster,
with the aim of obtaining both photometric and redshift data at
optical wavelengths and better X-ray data from the ROSAT data bank
(Pislar et al. 1997, Lima--Neto et al. 1997). We present here our photometric
data. The redshift catalogue is published in a companion paper (Durret
et al. 1997a) and the analysis of all these combined optical data will be
presented in Paper III (Durret et al. in preparation).

\section{The photographic plate data}

\subsection{Method for obtaining the catalogue}

We decided to obtain a photometric catalogue of the galaxies in the
direction of the Abell 85 cluster of galaxies by first processing the field
681 in the SRC-J Schmidt atlas. This blue glass copy plate (IIIaJ$+$GG385)
was investigated with the MAMA (Machine \`a Mesurer pour l'Astronomie)
facility located at the Centre d'Analyse des Images at the Observatoire de
Paris and operated by CNRS/INSU (Institut National des Sciences de l'Univers).
In order to also get information on the neighbouring galaxy distribution, the
central 5$^{\circ}\times$~5$^{\circ}$ area has been searched for objects
using the on-line mode with the 10~$\mu$m step size available at that time.
The involved algorithmic steps are well-known. They can be summarized as
follows~:
first a local background estimate and its variance are computed from
pixel values inside a 256~$\times$~256 window, then pixels with a
number of counts higher than the background value plus three times the
variance are flagged, which leads to define an object as a set of
connected flagged pixels; an overlapping zone of 512 pixels is used in
both directions for each individual scan. Although this method may appear
rather crude, its efficiency is nevertheless quite high for properly
detecting and measuring simple and isolated objects smaller than the
background scale. The region where ABCG~85 is located is not crowded by
stellar images ($b_{\rm II}\simeq-72^{\circ}$), so that most of the objects
larger than a few pixels can indeed be detected this way. 
The result was a list of more than 10$^5$ objects distributed over
the $\sim$~25 square degrees of the field bounded by 0$^{\rm
h}$31$^{\rm mn}$30.4$^{\rm s}$ $<\alpha<$ 0$^{\rm h}$53$^{\rm
mn}$10.6$^{\rm s}$ and $-$12$^{\circ}$18'19.43" $<\delta<$
$-$7$^{\circ}$05'13.88" (equinox 2000.0, as hereafter), with their
coordinates, their shape parameters (area, elliptical modelling) and
two flux descriptors (peak density, sum of background-subtracted
pixel values).

The astrometric reduction of the whole catalogue was performed with
respect to 91 stars of the PPM star catalogue (Roeser \& Bastian 1991)
spread over the field, using a 3$^{\rm rd}$-order polynomial fitting. The
residuals of the fit yielding the instrumental constants were smaller
than 0.25 arcsecond and the astrometry of our catalogue indeed appears 
to be very good, as confirmed by our multi-object fibre spectroscopy where
the galaxies were always found to be very close ($<$~2.0 arcseconds, i.e.
3 pixels) to the expected positions.

Since the required CCD observations were not available at that time, a
preliminary photometric calibration of these photographic data has
been done using galaxies with known total blue magnitude. The
magnitude of stars is certainly much easier to define, but such
high-surface brightness objects suffer from severe saturation effects
on Schmidt plates when they are bright enough to be included in
available photometric catalogues. So, 83 galaxies were selected from
the Lyon Extragalactic Database (LEDA) in order to compare their
magnitude to their measured blue flux. A small region around each of
these objects was scanned and this image has been used: i) to identify
the object among its neighbours within the coordinate list and ii) to
assess the quality of the flux value stored in the on-line catalogue
with respect to close, overlapping or merged objects. The 74 remaining
undisturbed objects identified with no ambiguity came from eight
different catalogues in the literature. Whatever the intrinsic
uncertainties about the integrated MAMA fluxes are, systematic
effects were found with respect to the parent catalogue in a flux
versus magnitude plot, as well as discrepancies for some objects
between the LEDA and the Centre de Donn\'ees Astronomiques de Strasbourg (CDS)
databases.  Consequently, three catalogues including 12 objects were
removed and the LEDA magnitude of 5 objects was replaced by a CDS
value which seems in better agreement with their aspect and with the overall
trend when compared to similar objects.  Later, 7 objects far from the overall
trend were discarded. These successive rejections resulted in a set of
55 objects distributed over a six magnitude range. The magnitude
zero-point for our photographic catalogue was obtained by plotting the
flux of these objects against their expected magnitude. A rms scatter
of 0.34 mag was computed around the linear fit.

\subsection{Classification of the objects}

Most of the diffuse objects included in our main catalogue were
automatically selected according to their lower surface brightness
when compared to stars.  As usual for glass copies of survey plates,
the discrimination power of this brightness criterion drops sharply
for objects fainter than approximately 19$^{\rm th}$ magnitude, and so does
the completeness of the resulting catalogue if no
contamination is allowed for. The number of galaxy candidates brighter
than this limit within the investigated area appeared, however, to be
already large enough to get a much better view of the bright galaxy
distribution than using the deeper but very incomplete catalogue
published by Murphy (1984). Moreover, including faintest objects was
not necessary for the redshift survey of the Abell~85 cluster of
galaxies we were planning (see Durret et al. 1997). Hence, no attempt was
done to reach a fainter completeness limit. Nonetheless, in order to
select galaxies, the decision curve which has been computed in the Flux
vs. Area parameter space was fitted to the data so that
some objects identified by Murphy from CCD frames as faint galaxies
were also classified as galaxies by us. Next, a further test based on
the elongation was performed in order to reject linear plate flaws or
artefacts, as well as to pick bright elongated galaxies first
classified as stars due to strong saturation effects. Finally,
spurious detections occuring around very bright stars (area greater
than 10$^3$ pixels) due to a wrong estimate of the local background were
tentatively removed by checking their location with respect to these
bright objects.
In this way, a list of more than 25,000 galaxy candidates over the 25
square degrees of our SRC-J~681 blue field was obtained.

\begin{figure*}[ht!]
\centerline{\psfig{figure=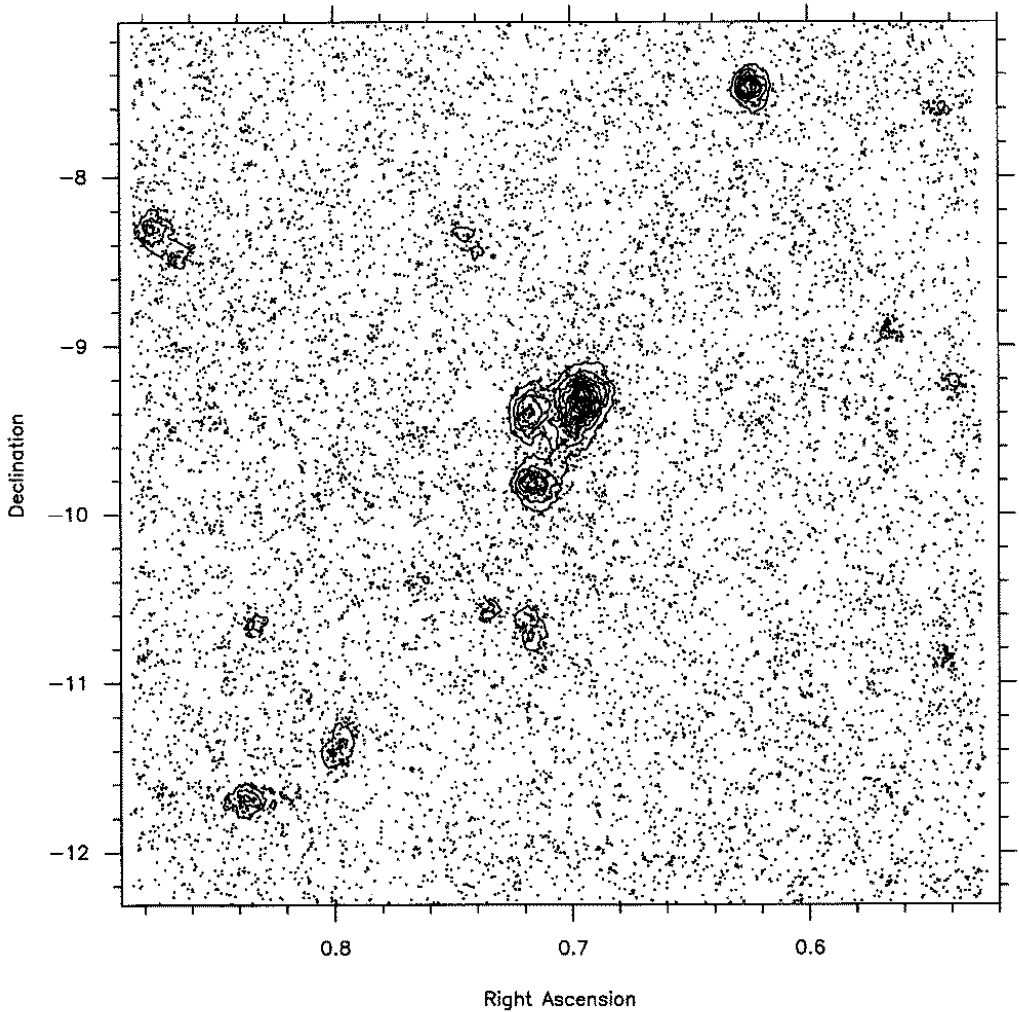,height=14cm}}
\caption[ ]{Spatial distribution of the 11,862 galaxies brighter than 
B$_{\rm J}=$~19.75 in the SRC-J~681 field. The large overdensities are 
indicated by 
superimposed isopleths from a density map computed by the method introduced 
by Dressler (1980) with $N=$~50~; eleven isopleths are drawn from 850 to 
2,850 galaxies/square degree.}
\protect\label{allplate}
\end{figure*}

The distribution of these galaxies is displayed in Fig.~\ref{allplate}
for objects brighter than B$_{\rm J}=$~19.75. The Abell~85 cluster is
clearly visible, as well as several other density enhancements which
are mostly located along the direction defined by the cluster
ellipticity.

\subsection{Completeness and accuracy of the classification}

\begin{figure}[h!]
\centerline{\psfig{figure=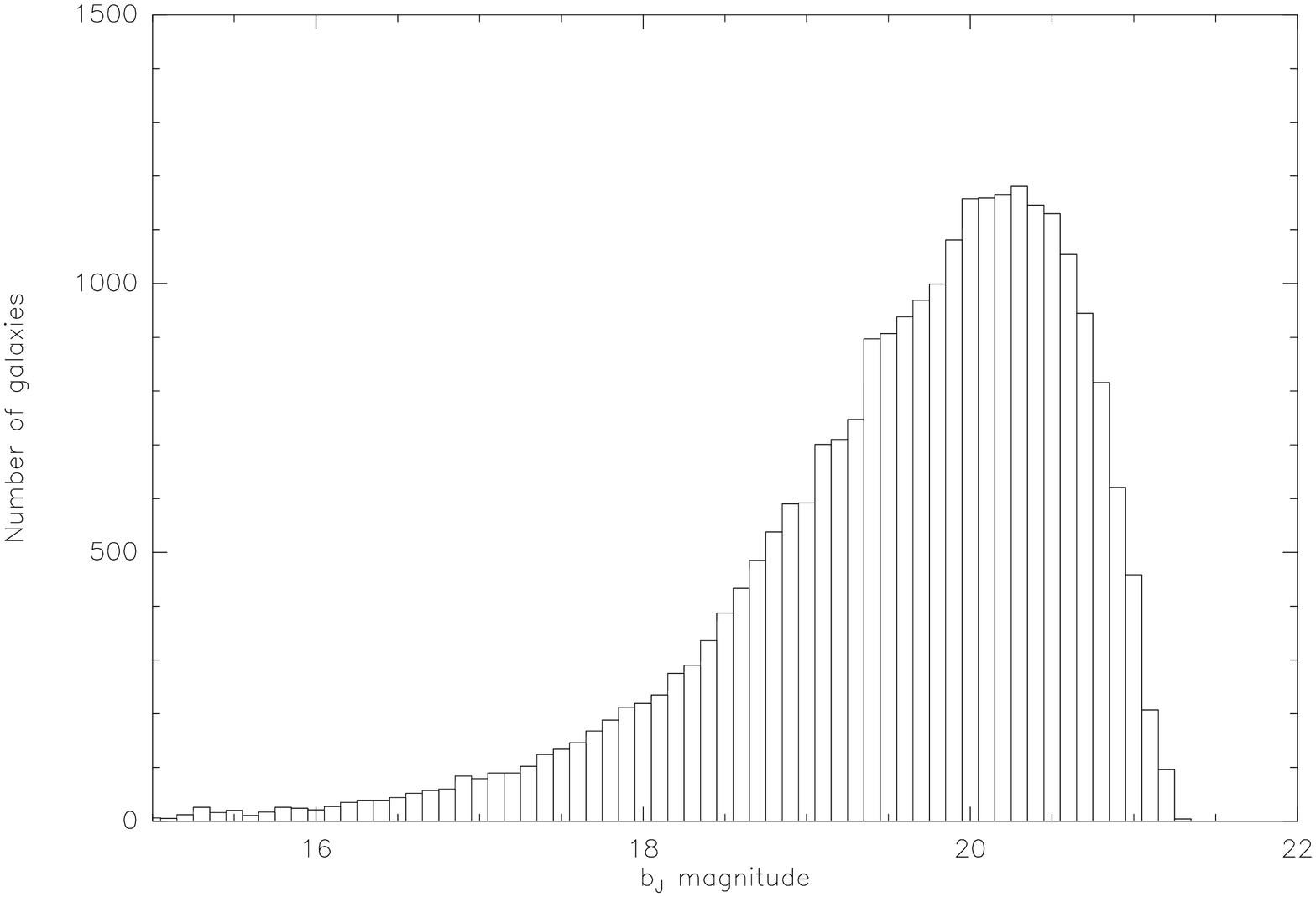,height=7cm,angle=-90}}
\caption[ ]{Differential magnitude distribution of the 25~10$^3$ galaxy 
candidates in the SRC-J 681 field.}
\protect\label{fdl}
\end{figure}

The differential luminosity distribution of the galaxy candidates
indicates that the sample appears quite complete down to the
B$_{\rm J}=$~19.75 magnitude (see Fig.~\ref{fdl}).  To go further, we first
tested the completeness of this overall list by cross-identifying it
with three catalogues from the literature (Murphy 1984, Beers et
al. 1991, Malumuth et al. 1992) with the help of images obtained from
the mapping mode of the MAMA machine. It appeared that: i) all but one
galaxy of the Malumuth et al. (1992) catalogue of 165 objects are actually
classified as galaxies, with a mean offset between individual
positions equal to 1.10~$\pm$~0.06 arcsecond ; ii) 94\% of the 35
galaxies listed by Beers et al. (1991) inside the area are included in our
catalogue, only 2 bright objects which suffer from severe saturation
being misclassified. Note that such an effect also caused 5 of the 83
galaxies chosen as photometric standards to be misclassified, which
gives the same percentage as for the sample by Beers et al. The
comparison with the faint CCD catalogue built by Murphy (1984) in the so-called
$r_{\rm F}$ band (quite similar to that obtained using a photographic IIIaF
emulsion with a R filter) was performed only for objects which were visible
on the photographic plate with secure identification (only uncertain X
and Y coordinates are provided in the paper) and classified without any
doubt as galaxies from our visual examination. There remained 107
objects out of 170, among which 88 are brighter than B$_{\rm J}\sim$~19.75
($r_{\rm F}\sim$~18.5).  Down to this flux limit, 82 objects ($\sim$~93\%)
are in agreement, thereby validating the choice of our decision curve in the
Flux vs. Area parameter space. These cross-identifications therefore indicate
that the completeness limit of our catalogue is about 95\% for such objects,
as expected from similar studies at high galactic latitude.

In order to confirm this statement and to study the homogeneity of our
galaxy catalogue, we then decided to verify carefully its reliability
inside the region of the Abell~85 cluster of galaxies itself. The
centre of ABCG~85 was assumed to be located at the equatorial
coordinates given in the literature, $\alpha=$~0$^{\rm h}$41$^{\rm
mn}$49.8$^{\rm s}$ and $\delta=$~$-$9$^{\circ}$17'33.", and a square
region of $\pm$1$^\circ$ around this position was defined; such an
angular distance corresponds to $\sim$~2.7~Mpc~h$_{100}^{-1}$ at the
redshift of the cluster ($z=$~0.0555). However, let us remark that the
position of the central D galaxy is slightly different,
$\alpha=$~0$^{\rm h}$41$^{\rm mn}$50.5$^{\rm s}$ and
$\delta=$~$-$9$^{\circ}$18'11.", and so is the centre we found from
our X-ray analysis of the diffuse component of this cluster, i.e.:
$\alpha=$~0$^{\rm h}$41$^{\rm mn}$51.9$^{\rm s}$ and
$\delta=$~$-$9$^{\circ}$18'17." (Pislar et al. 1997). For all our
future studies, we then chose to define the cluster centre as that of
this X-ray component.

The distribution of the $\sim$~4,100 candidates within the area has
been first of all visually inspected to remove remaining conspicuous
false detections around some stars as well as some defects mainly due to a
satellite track crossing the field. This cleaned catalogue contains a
little more than 4,000 galaxy-like objects, half of which brighter
than B$_{\rm J}=$~19.75.  The intrinsic quality of this list has then been
checked against a visual classification of all the recorded objects
within a $\pm$~11'25" area covering the region already
observed by Murphy (1984) around the location $\alpha=$~0$^{\rm
h}$41$^{\rm mn}$57.0$^{\rm s}$ and $\delta=$~$-$9$^{\circ}$23'05". The
inspection of the corresponding MAMA frame of 2048~$\times$~2048
pixels enabled us to give a morphological code to each object, as well
as to flag superimposed objects and to deblend manually 10 galaxies
(new positions and flux estimates for each galaxy member). Of course,
the discrimination power of this visual examination decreases for
star-like objects fainter than B$_{\rm J}=$~18.5 ($r_{\rm F}\sim$~17.3)
due to the sampling involved (pixel size of 0.67"), and an exact
classification of such objects appeared to be hopeless above the {\it
a priori} completeness limit of our automated galaxy list guessed to
be B$_{\rm J}=$~19.75. Down to this limit, our results can be summarized as
follows~: i) $\sim$94\% of the selected galaxies are true galaxies
(including 7 multiple galaxies and 2 mergers with stars), while 4\%
may be galaxies~; ii) 7 genuine galaxies are missed (4\%).  Since these
contamination and incompleteness levels of 5--6\% were
satisfactory, we decided to set the completeness limit for our
automated galaxy catalogue at this magnitude B$_{\rm J}=$~19.75.

\subsection{The photographic plate catalogue}

\begin{figure}[ht!]
\centerline{\psfig{figure=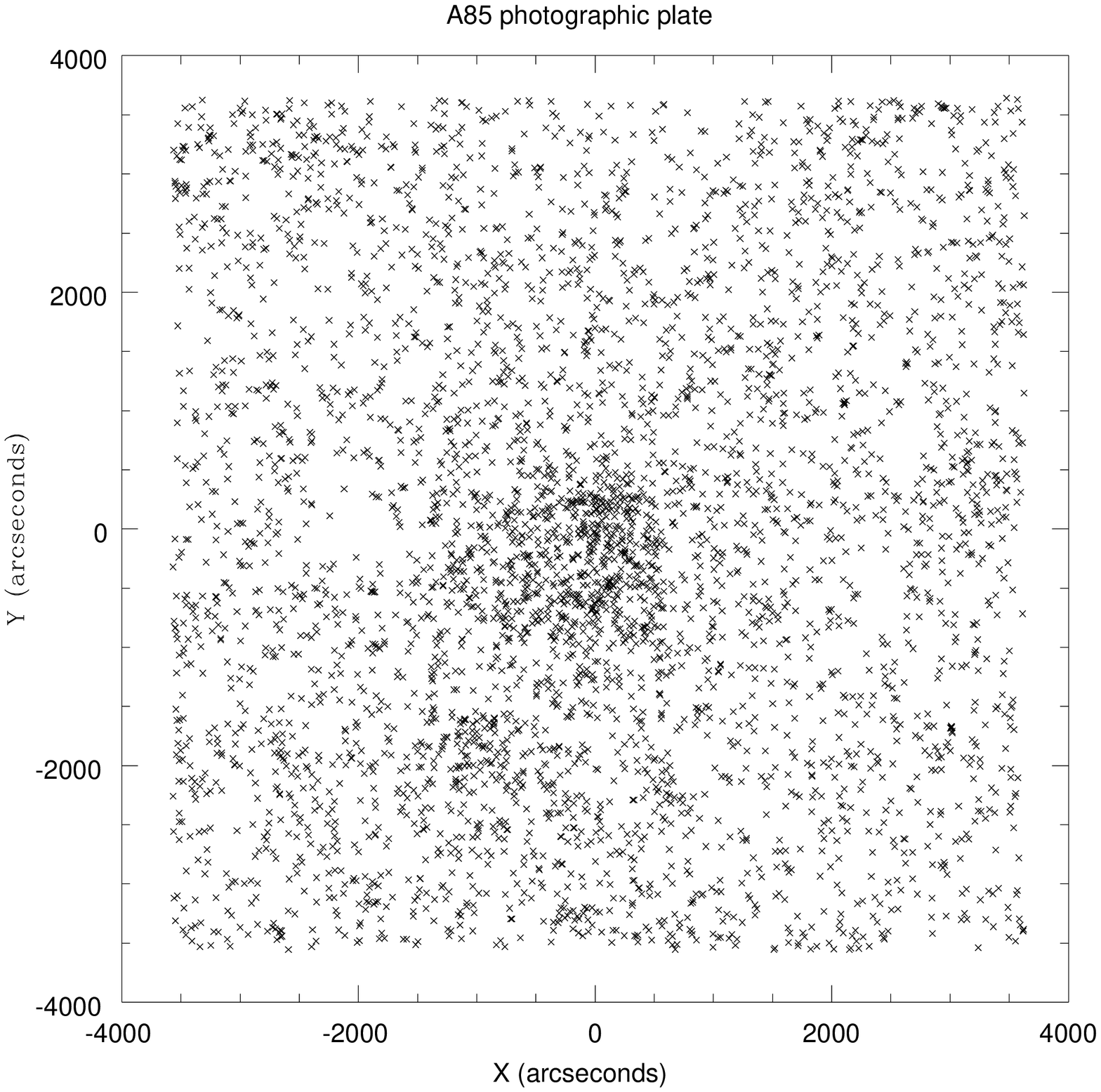,height=8cm}}
\caption[ ]{Positions of the 4,232 galaxies detected on the photographic plate
relative to the centre of the cluster defined as the centre of the diffuse
X-ray component. North is to the top and East to the left.}
\protect\label{mamaxy}
\end{figure}

For objects fainter than our completeness limit, the visual check of
the inner ($\pm$~11'25") part of our object list has enabled us to
confirm the galaxy identification of 135 galaxy candidates as well as
to select 214 misclassified faint galaxies.  The total number of
galaxies included in the visual sample down to the detection limit is
541, whereas the initial list only contains 338 candidates within the
same area. Keeping in mind that both catalogues are almost identical
for objects brighter than B$_{\rm J}=$~19.75, we decided to replace the
automated list by the visual one inside this $\pm$~11'25" central
area. Note that about 150 objects remained unclassified, including 26
galaxies from the CCD list by Murphy. We added these 26 galaxies to
the final catalogue whose galaxies are plotted in Fig.~\ref{mamaxy}.

Table~1 lists the merged catalogue of 4,232 galaxies obtained from the 
SRC-J~681 plate in the $\pm 1^{\circ}$ field of ABCG~85, with V and R
magnitudes computed using the transformation laws obtained from our CCD
data (see \S 3.3). This Table includes the following information~: running
number~; equatorial coordinates (equinox 2000.0)~; ellipticity ; position
angle of the major axis~; B$_{\rm J}$, V, and R magnitudes~; X and Y
positions in arcsecond relative to the centre defined as that of the diffuse
X-ray emission of the cluster (see above)~; cross-identifications with the
lists by Malumuth et al. (1992), Beers et al. (1991) and Murphy (1984).

\section{The CCD data}

\subsection{Description of the observations}

\begin{figure}[ht!]
\centerline{\psfig{figure=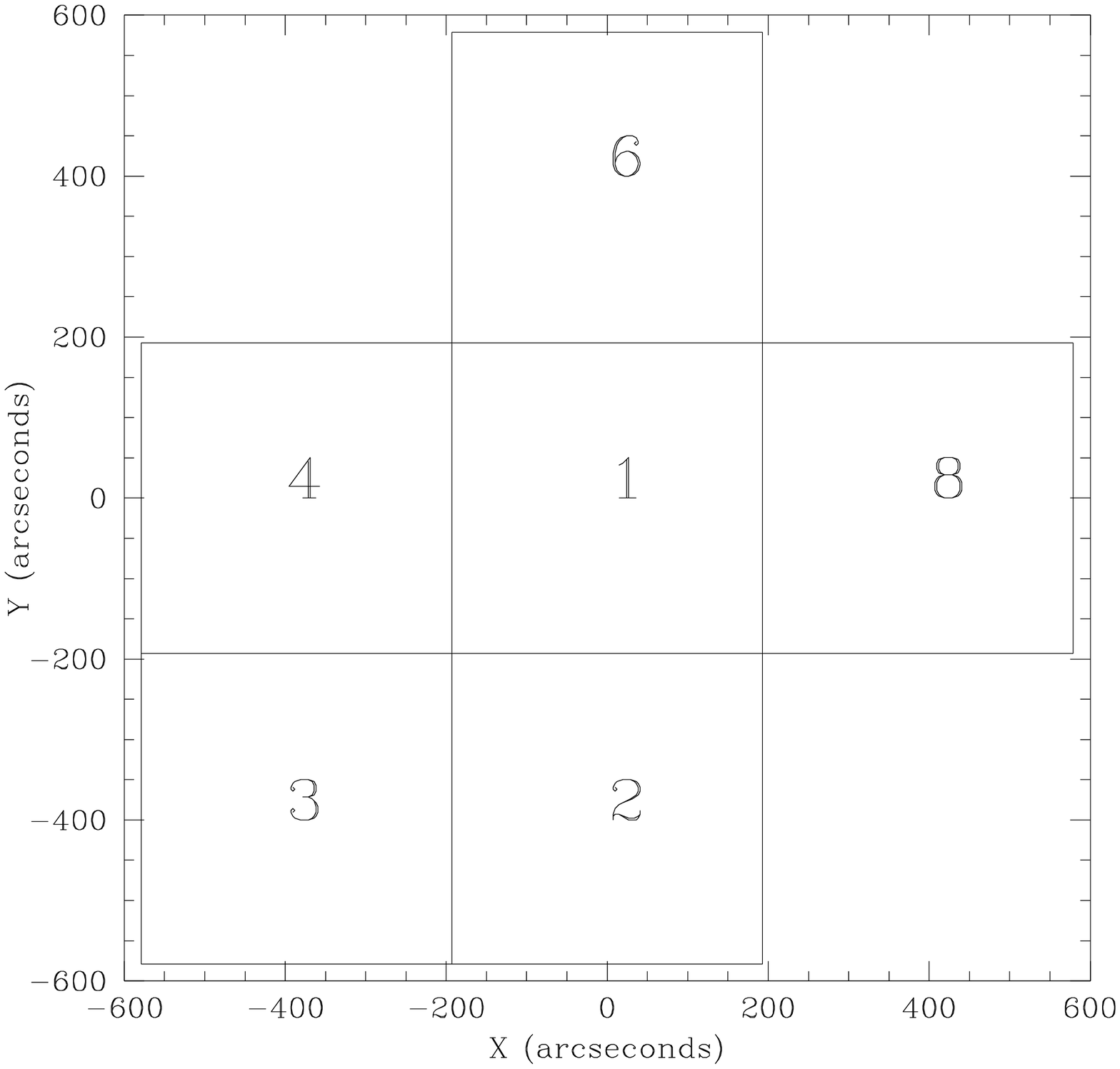,height=5cm}}
\caption[ ]{Distribution of the fields observed with CCD imaging. The size
of each field is 6.4$\times$6.4~arcmin$^2$. Positions are drawn relatively to 
the centre with equatorial coordinates
$\alpha=$~0$^{\rm h}$41$^{\rm mn}$46.0$^{\rm s}$ and
$\delta=$~$-$9$^{\circ}$20'10".}
\protect\label{champsccd}
\end{figure}

The observations were performed with the Danish 1.5m telescope at
ESO La Silla during 2 nights on November 2 and 3, 1994 (the third night was 
cloudy, and this accounts for the missing fields in Fig.~\ref{champsccd}). 
A sketch of the observed fields is displayed in Fig.~\ref{champsccd}. Field~1
was centered on the coordinates~: 00$^{\rm h}$41$^{\rm mn}$46.00$^{\rm s}$, 
$-9^\circ$20'10.0" (2000.0). There was almost no overlap between the
various fields (only a few arcseconds). The Johnson
V and R filters were used. Exposure times were 10~mn for all fields; 1~mn 
exposures were also taken for a number of fields with bright objects in
order to avoid saturation. The detector was CCD~\#28 with 1024$^2$ pixels of 
24~$\mu$m, giving a sampling on the sky of 0.377"/pixel, and a size of 
6.4$\times$6.4~arcmin$^2$ for each field. 
The seeing was poor the first night~: 1.5--2" for fields 1 and 2, 2--3" for 
field 3 (in which consequently the number of galaxies detected is much 
smaller), and good the second night~: 0.75--1.1". On the 
other hand, the photometric quality of the first night was better than that
of the second one. However, the observation of many standard stars per night
made a correct photometric calibration possible even for the second night as
indicated by a comparison with an external magnitude list~: the photometric
catalogues from the six fields have the same behaviour for both nights
(see e.g. Fig.~8).

\subsection{Data reduction}

Corrections for bias and flat-field were performed in the usual way with
the IRAF software. Only flat fields obtained on the sky at twilight and
dawn were used; dome flat fields were discarded because they showed too
much structure.

Each field was reduced separately.  The photometric calibration took
into account the exposure time, the time at which the exposure had
been made, the color index (V-R), the airmass, and a second order term
including both the color index and airmass. The photometric
characteristics of both nights were estimated separately.

Objects were automatically detected using the task
DAOPHOT/DAOFIND. This task performs a convolution with a gaussian
having previously chosen characteristics, taking into account the
seeing in each frame (FWHM of the star-like profiles in the image) as
well as the CCD readout noise and gain. Objects are identified as the
peaks of the convolved image which are higher than a given threshold
above the local sky background (chosen as approximately equal to
4~$\sigma$ of the image mean sky level).  A list of detected objects
is thus produced and interactively corrected on the displayed image so
as to discard spurious objects, add undetected ones (usually close to
the CCD edges) and dispose of false detections caused by the events
flagged in the previous section. Since exposure times were the same in
V and R, the number of objects detected in the R band is of course much
larger.

\begin{figure}
\centerline{\psfig{figure=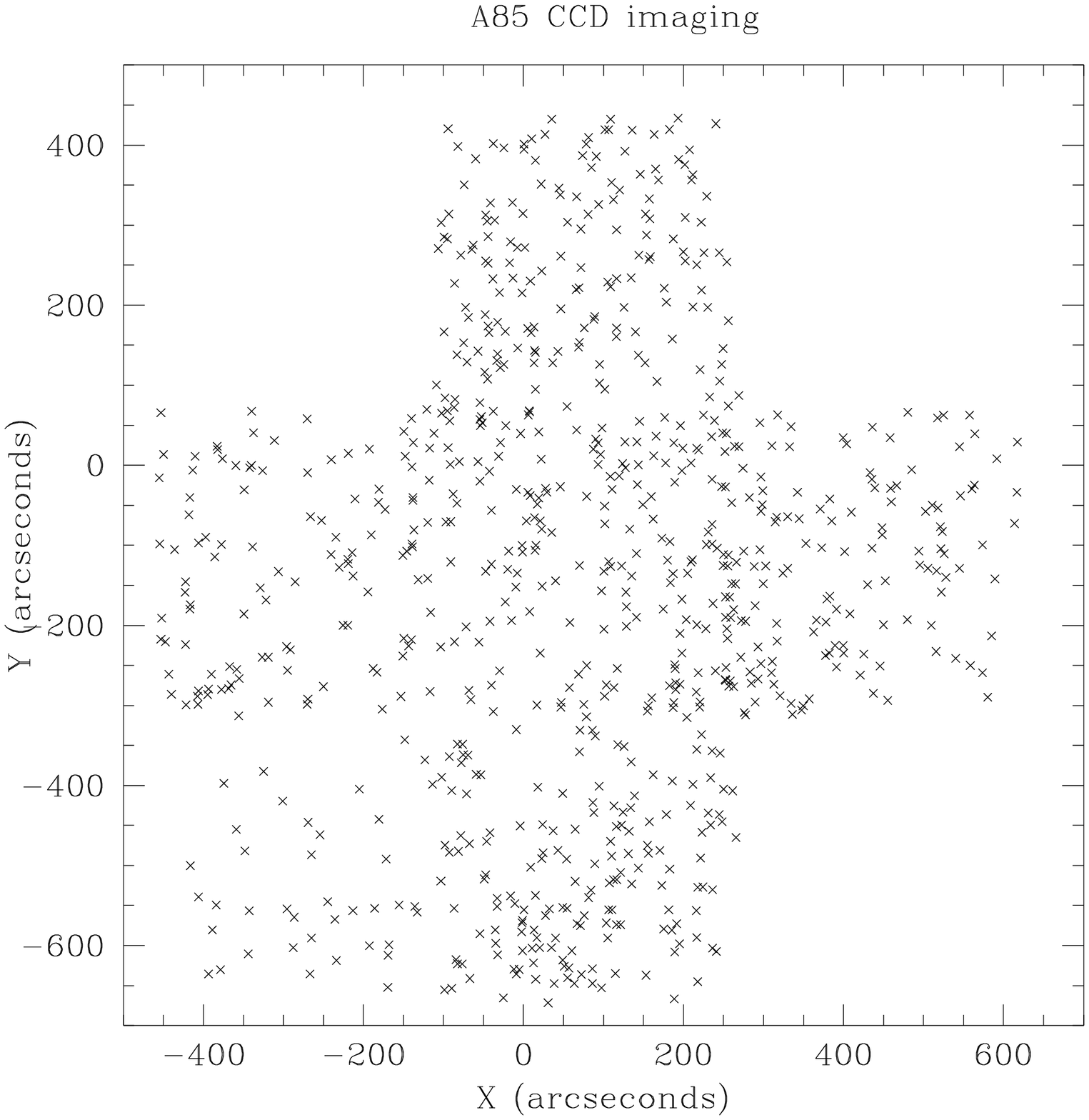,height=6cm}}
\caption[ ]{Positions of the galaxies detected in the R band relative to the
centre defined as the centre of the diffuse X-ray emission (see text).}
\protect\label{ccdxy}
\end{figure}

We used the package developed by O.~Le F\`evre (Le F\`evre et al. 
1986) to obtain for each field a catalogue with the (x,y) galaxy positions, 
isophotal radii, ellipticities, major axis, position angles, and V and R 
magnitudes within the 26.5 isophote.
Star-galaxy separation was performed based on a compactness parameter q
determined by Le F\` evre et al. (1986, see also Slezak et al. 1988), as
described in detail e.g. by Lobo et al. (1997). We chose q=1.45 as the best
separation limit between galaxies and stars; very bright stars were 
classified as galaxies with this criterion and had to be eliminated manually.
After eliminating repeated detections of a few objects, we obtained a total 
number of 805 galaxies detected in R, among which 381 are detected in V. 
The errors on these CCD magnitudes are in all cases smaller than 0.2 magnitude,
and their rms accuracy is about 0.1 magnitude; these rather large values are
due to the bad seeing during the first night and to pretty poor photometric
conditions during the second night.

Positions of the galaxies detected in the R band relative to the
centre defined above are displayed in Fig.~\ref{ccdxy}. Notice the smaller
number of galaxies detected in field 3 due to a sudden worsening of the seeing 
during the exposure on this field.
The astrometry of this CCD catalogue is accurate to 1.5--2.0 arcseconds as
verified from the average mutual angular distance between CCD and MAMA
equatorial coordinates for 174 galaxies included in both catalogues.

\begin{figure}
\centerline{\psfig{figure=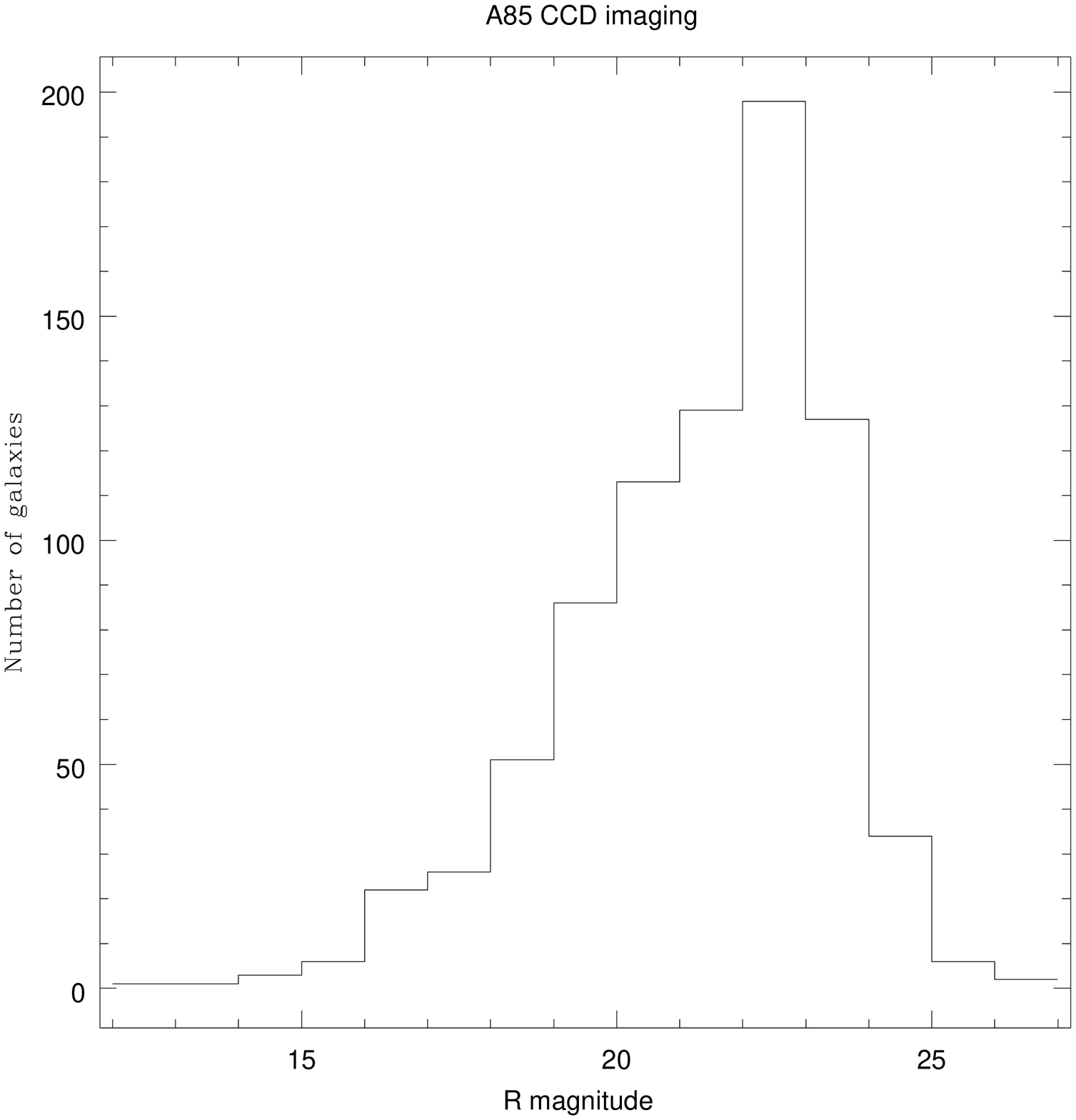,height=7cm}}
\caption[ ]{Histogram of all the R magnitudes of the galaxies in the
CCD catalogue.}
\protect\label{ccdrmag}
\end{figure}

The histogram of the R magnitudes in the CCD catalogue is displayed in
Fig.~\ref{ccdrmag}. It will be discussed in detail in Paper~III
(Durret et al.  in preparation). The turnover value of this histogram
is located between R=22 and R=23, suggesting that our catalogue is
roughly complete up to R=22.

\begin{figure}
\centerline{\psfig{figure=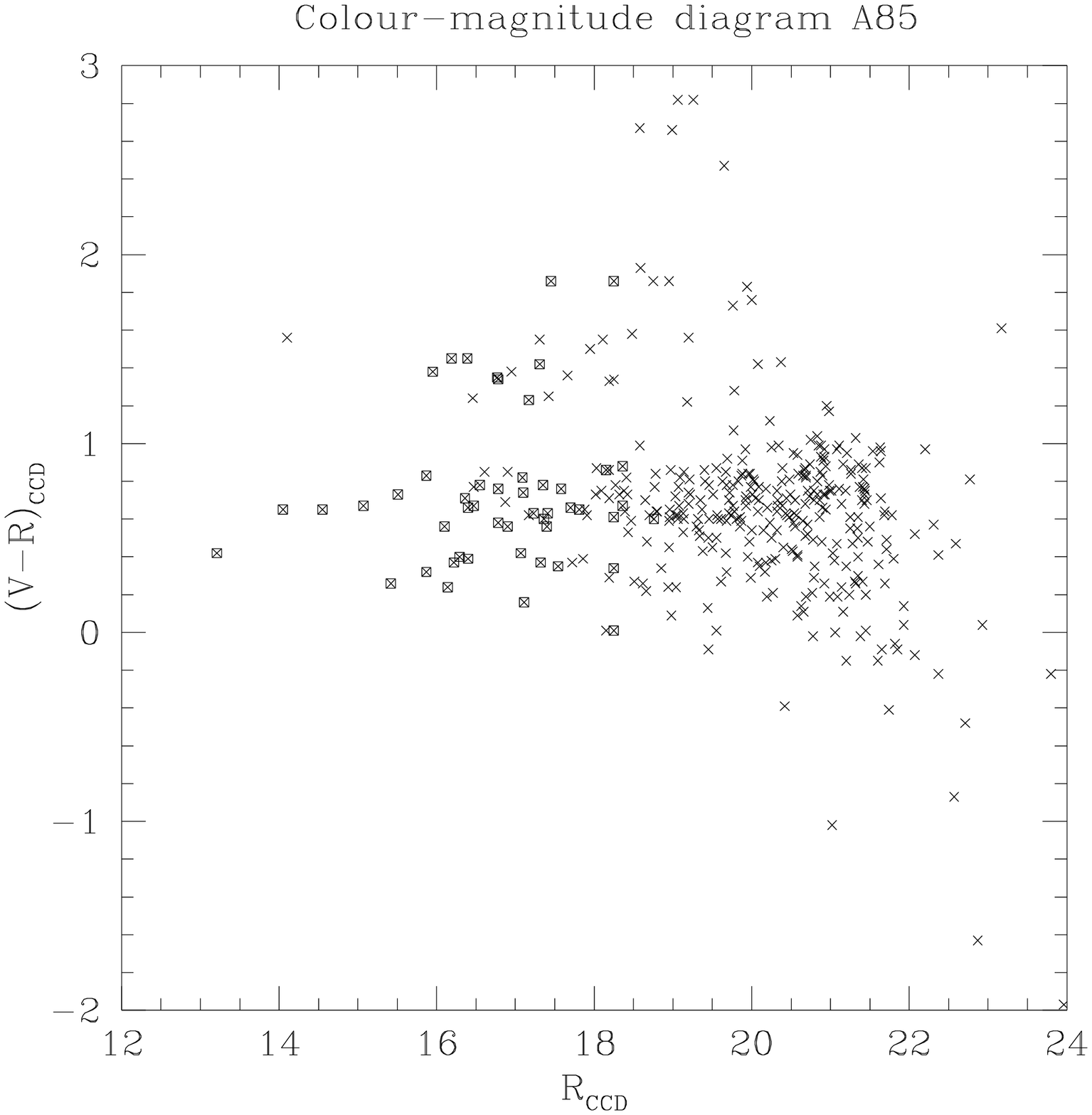,height=8cm}}
\caption[ ]{(V-R) colour as a function of R for the 381 galaxies detected
in the V band in our CCD catalogue. The 50 galaxies indicated with a square are
those with redshifts in the interval 13,350 -- 20,000~\kms assumed to belong to
\a85.}
\protect\label{coul}
\end{figure}

The (V-R) colours are plotted as a function of R for the 381 galaxies
detected in the V band in our CCD catalogue
(Fig.~\ref{coul}). Unfortunately, since the observed CCD field is
small, there are only 50 of these galaxies with measured redshifts,
and therefore it is not possible to derive a colour-magnitude relation
from which to establish a membership criterion for the cluster.

\subsection{Transformation laws between the photometric systems}

576 stars were also measured on the CCD images and used to calculate 
calibration relations between our photographic plate \bj\ magnitudes and 
our V and R CCD magnitudes.

For stars:\\
$$V_{CCD}=B_{\rm J}- 40.8302+3.6656\ B_{\rm J} -0.082567\ B_{\rm J}^2 $$
$$R_{CCD}=B_{\rm J} -10.12663+0.430772\ B_{\rm J} $$

For galaxies where only R is detected:\\
$$R_{CCD}=B_{\rm J} -3.03532+0.121963\ B_{\rm J}$$

For galaxies where both V and R are detected:\\
$$V_{CCD}=B_{\rm J}-2.13942+0.108905\ B_{\rm J}$$
$$R_{CCD}=B_{\rm J}-0.566762(V-R)-2.29919+0.10482\ B_{\rm J}$$

\begin{figure}
\centerline{\psfig{figure=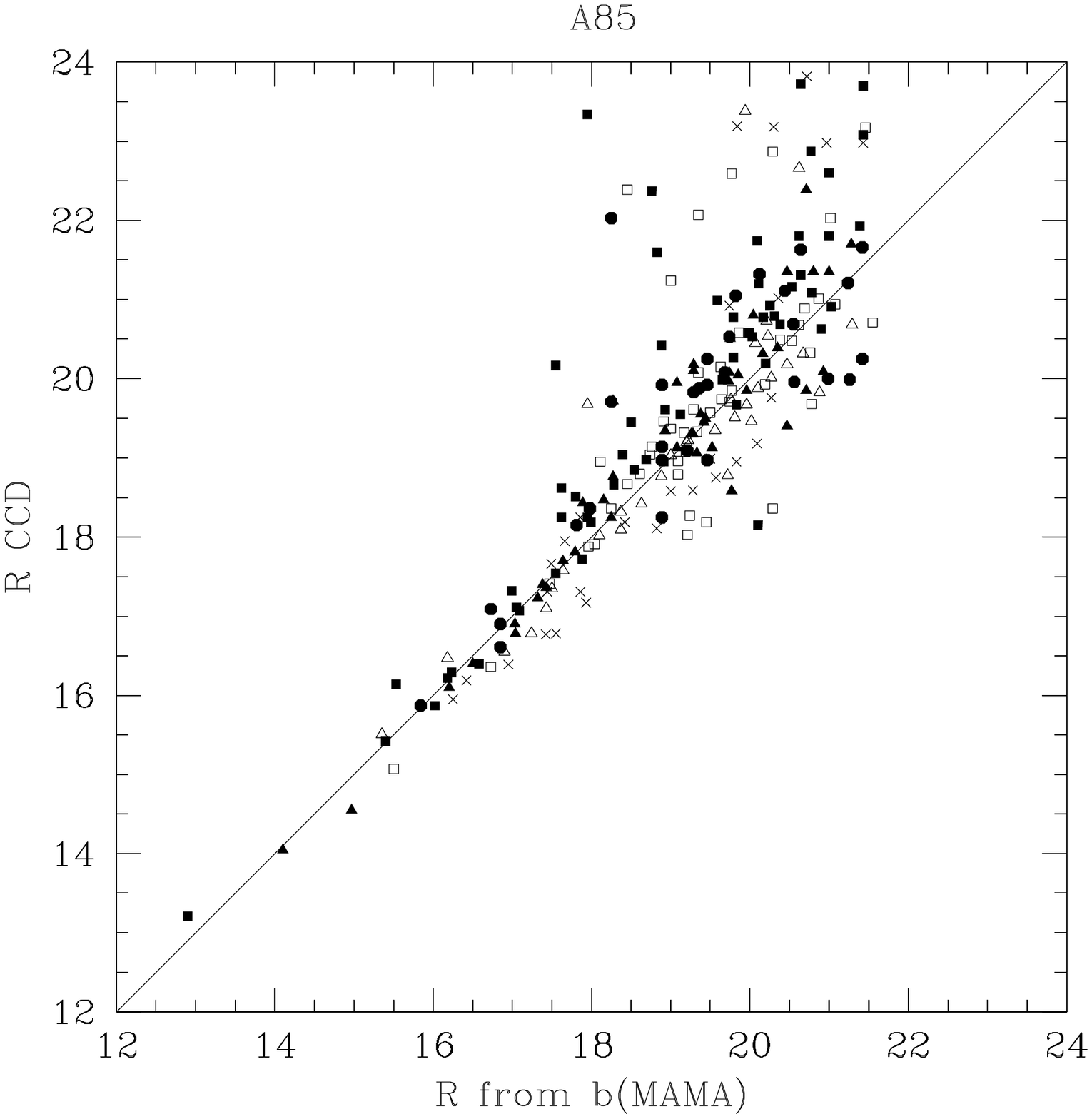,height=8cm}}
\caption[ ]{Observed R band CCD magnitude $R_{CCD}$ as a function of the 
R magnitude calculated from the photographic \bj magnitude. The six different 
symbols correspond to the six CCD fields described above.}
\protect\label{rr5}
\end{figure}

The observed R band CCD magnitude $R_{CCD}$ as a function of the R
magnitude calculated from the photographic \bj magnitude is plotted
in Fig.~\ref{rr5} for galaxies, showing the quality of the correlation
for the six different CCD fields, especially for objects brighter than R$=$19.
All the CCD fields appear to behave identically.

\subsection{The CCD catalogue}

The CCD photometric data for the galaxies in the field of \a85 are given 
in Table~2. This Table includes for each object the following information~:
running number~; equatorial coordinates (equinox 2000.0)~; isophotal radius~;
ellipticity~; position angle of the major axis~; V and R magnitudes~; X and
Y positions in arcsecond relative to the centre assumed to have coordinates
$\alpha = 0^{\rm h}41^{\rm mn}51.90^{\rm s}$ and 
$\delta = -9^\circ$18'17.0" (equinox 2000.0) (this centre was chosen to
coincide with that of the diffuse X-ray gas component as defined by Pislar
et al. (1997) ).\\

\section{Conclusions} 

Our redshift catalogue is submitted jointly in a companion paper
(Durret et al. 1997). Together with the catalogues presented here, it
is used to give an interpretation of the optical properties of \a85
(Durret et al. in preparation, Paper~III), in relation with the X-ray
properties of this cluster (Pislar et al. 1997, Lima--Neto et al. 1997, 
Papers~I and II).

\acknowledgements {We are very grateful to the MAMA team at
Observatoire de Paris for help when scanning the photographic plate,
and to Cl\'audia Mendes de Oliveira for her cheerful assistance at the
telescope.  CL is fully supported by the BD/2772/93RM grant attributed
by JNICT, Portugal.}

\end{document}